# Curating Transient Population in Urban Dynamics System

Gautam S. Thakur[1], Kevin A. Sparks[1], Robert N. Stewart[1],
Marie L. Urban[1], Budhendra L. Bhaduri[1]

[1]The Geographic Information Science and Technology (GIST) Group
Oak Ridge National Laboratory, 1 Bethel Valley Road, Oak Ridge, TN 37831 USA
Email: {thakurg, sparksa, stewartrn, urbanml, bhaduribl}@ornl.gov

## Abstract

For past several decades, research efforts in population modelling has proven its efficacy in understanding the basic information about residential and commercial areas, as well as for the purposes of planning, development and improvement of the community as an eco-system. More or less, such efforts assume static nature of population distribution, in turn limited by the current ability to capture the dynamics of population change at a finer resolution of space and time. Fast forward today, more and more people are becoming mobile, traveling across borders impacting the nuts and bolts of our urban fabric. Unfortunately, our current efforts are being surpassed by the need to capture such transient population. It is becoming imperative to identify and define them, as well as measure their dynamics and interconnectedness. In this work, we intend to research urban population mobility patterns, gauge their transient nature, and extend our knowledge of their visited locations. We plan to achieve this by designing and developing novel methods and using VGI data that models and characterizes transient population dynamics.

## 1. Introduction

Over the last decade, worldwide percentages of transient population i.e. population on-the-move have grown linearly, on account of socio-economic development and rising global conflicts (Worldbank 2016). Characterizing such movement patterns are vital to understand the dynamics of an urban system. Beyond that, privacy-preserved and discrete insight into the diurnal activities of people, where they are, where they go, and how much time they spent at certain locations is vital for first responders in times of emergencies, and for urban planners to develop robust future cities. Traditionally, approaches involved in curating such information have heavily relied on census data, surveys, and simulations models. While they provide an excellent source of baseline statistics - for logistical purposes, they rarely collect the spatio-temporal distribution relating to the dynamics of population movements. Furthermore, they omit non-residential locations, such as business districts, parks, and museums. Social media(Hawelka et al. 2014; Frias-Martinez et al. 2012), cellular data(Di Lorenzo & Calabrese 2011; Calabrese et al. 2010; Calabrese et al. 2011; Calabrese et al. 2013; Isaacman et al. 2012) have proved their efficacies in representing global patterns of human mobility. However, much of the current research has not focused on curating dynamics of population. This work attempts to provide a set of guidelines that help infer the rise of transitivity patterns and generate population numbers. In this work, first we propose *Transient Population Dynamics Model* that attempts to explain the mobility patterns of active population. In addition, it provides approaches to identify and estimate transient population and aid in discovering new locations and the duration of their visits. Second, we utilize this model in estimating the transient population across Australia and compare the differences against respective census data. Also, we intend to discover frequently visited locations in outback and metropolitan areas. We believe this work will generate enthusiasm among researchers and make a case for the importance of curating transient population.

## 4. Model for Transient Population

The Transient Population Model (TPM) attempts to provide insight into the dynamics of an urban system by estimating population mobility and activities governing them. The purpose of the model is to - i) Identify transient population; ii) Estimate transient population; iii) Discover locations (facilities) containing transient population; and iv) Estimate duration of transient activity. For a given geographical region with population $N= \{1,2,3...,n\}$, we define the subset of this population $\tilde{N}=\{1,2,3,...,\tilde{n}\}$ as transient; total active = $\eta$, such that $N \geq \tilde{N}$, moving among a set of locations $L=\{1,2,3...,l\}$ (total of $\Gamma$ locations) in that region at any given time. An individual $\tilde{N}_i$ is associated with its base location(e.g. home) is transient, if moving from location $l_j \rightarrow l_k$ or already moved to a new location $l_k$.

### 4.1 Identify transient population:

The transient population is identified based on the movement patterns between a pair of locations or while away from their base location. Thus, person $\tilde{N}_i$ is transient when,

$$T(\tilde{N}_i) \rightarrow \begin{cases} 1; & if\ l_j \rightarrow l_k \\ 1; & if\ \tilde{N}_i\ is\ at\ location\ l_k \\ 0; & otherwise \end{cases}$$

(1)

Where $T(.)$ is the transient function, representing the movement of $\tilde{N}_i$.

### 4.2 Estimate transient population

The aggregation of movements of individual $\tilde{N}_i$ over an observed time provides the estimate of its mobility patterns in the given geographical region. The total number of movement patterns of $\tilde{N}_i$ over a time interval $t$ is given by,

$$M(\tilde{N}_i) = \beta \sum_{s=0;j,k=1;j\neq k}^{\Gamma} (\tilde{N}_i)_{t_s}^{l_j \rightarrow l_k}$$

(2)

for $M$ is the movement pattern profile for $\tilde{N}_i$ and is $\beta$ constant to reduce ping-pong effect (micro and short-term mobility)(Chiou 2007). Thus, cumulative transient estimates *(W)* can be calculated as the agglomeration of all $\tilde{N}_i$ over the time $t$.

$$W(M(\tilde{N}_i)) = \beta \sum_{i}^{\eta} \sum_{s=0;j,k=1;j\neq k}^{\Gamma} (\tilde{N}_i)_{t_s}^{l_j \rightarrow l_k}$$

(3)

### 4.3 Discover locations containing transient population

Here we discover a set of destination locations where transient population appears over time (exclude base/home location of the population). The motivation to identify such locations provide an insight into the activity patterns, which are otherwise not captured through census survey. In order to differentiate from the base location, we make an explicit assumption that transient location is characterized by a greater number of unique visitors. The agglomeration (*Q*) of transient population for a location $l_k$

$$Q(l_k) = \sum_{i=0;j=0,j\neq k}^{\eta,\Gamma} (\tilde{N}_i)_{t_s}^{l_j \rightarrow l_k}$$

(4)

## 4.4 Estimate duration of transient activity

The duration of transient activity is the sum of the time duration of movement of individuals in transient. For simplicity, we aggregate the total amount of time individuals' movements

$$\Theta(\tilde{N}) = \sum_{t} \sum_{;j,k=1;j \neq k}^{\Gamma} (\tilde{N}_i)_{t_s}^{l_j \rightarrow l_k}$$

(5)

## 5. Results and Analysis

We used the proposed model and data from several social media sources to estimate transient population and discover locations (facilities) in Australia.

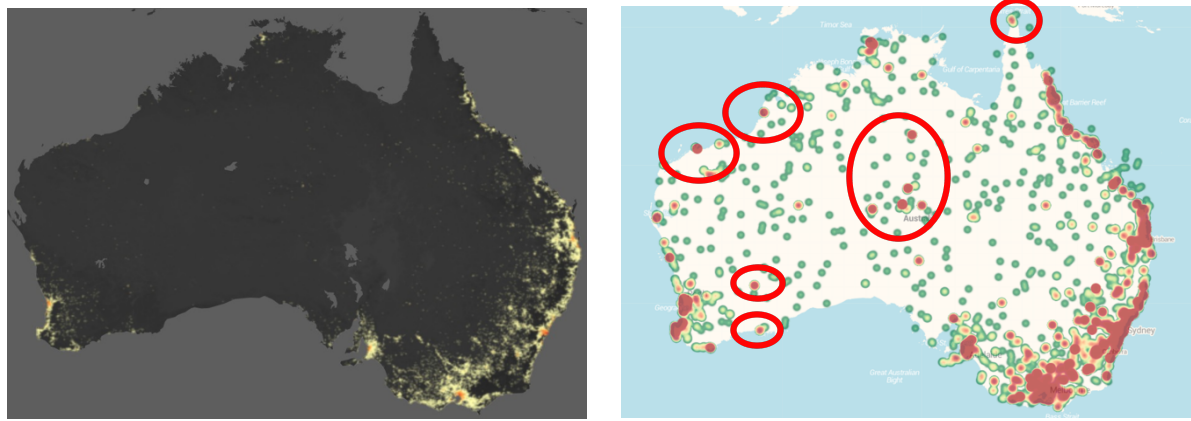

**Figure 1: Comparing the Census (a) vs. transient population (b) in Australia**

We compare our aggregate results with the census of Australia in Figure 1. In (b) we have circled out regions containing transient population, that is missing in the census map. This approach can be very helpful in uncovering regions which are otherwise not represented using traditional techniques (e.g. census and surveys).

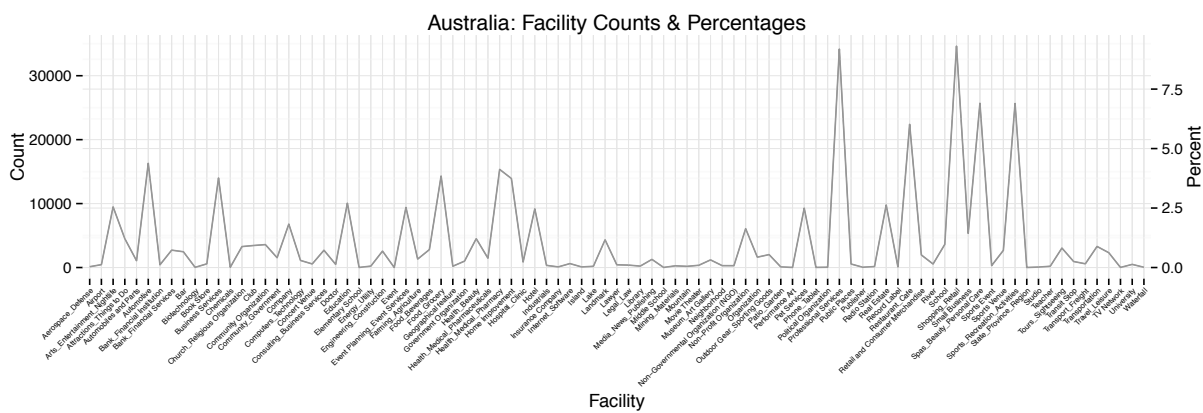

**Figure 2: Transient locations discovered in Australia.**

We explore each of these regions to generate a high resolution dataset of locations and population. In Figure 2, we show a list of locations and their frequency in Australia, which are mostly commercial and have people actively visiting them. In all, we have discovered more than 100k unique facilities in Australia and clustered them in 50+ categories.

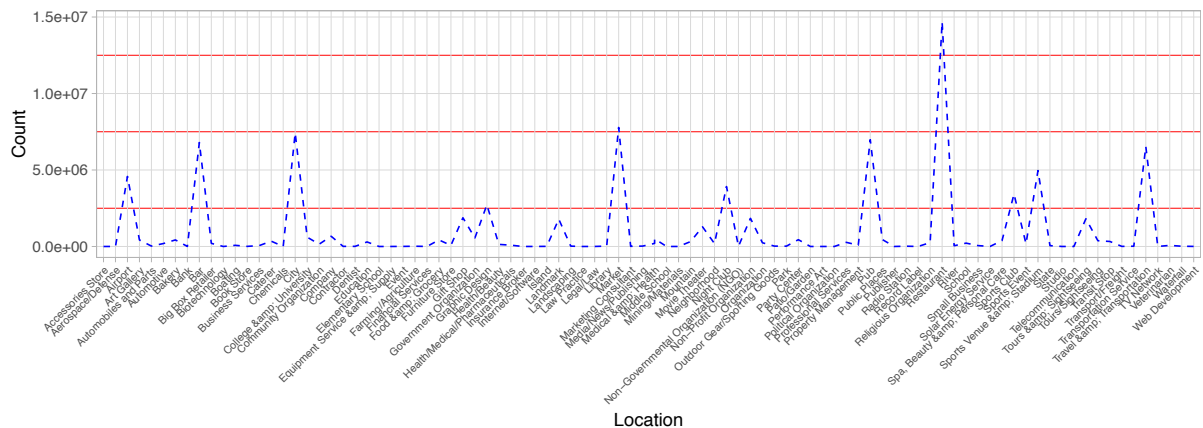

**Figure 3: Transient population in Australia.**

Figure 3 shows locations and approximate number of people for Australia. 'Restaurants', 'Market Places', and 'Pubs' have recorded the highest number of active population visits during the observed time period. Thus the proposed work also provides insights into ranking location types based on transient population.

## 6. Conclusion

This research provided an insight into the nature of transient population – their dynamics, mobility patterns, and discovering locations where such population might trend. In addition, we proposed a model to discover and quantify population's structure and emergence. We used VGI data from several social media sources to demonstrate our approach for studying transient population and emerging locations in Australia. In future, we would like to expand on this idea, by first improving the temporal signatures at even finer resolution (of day and hours), and second, to capture seasonal variations appearing over the years. We hope our work will provide a more realistic and accurate approach for generating population signatures and to aid in the analysis, simulation, and design of future urban systems.

## Acknowledgements

This manuscript has been authored by UT-Battelle, LLC under Contract No. DE-AC05-00OR22725 with the U.S. Department of Energy. The United States Government retains and the publisher, by accepting the article for publication, acknowledges that the United States Government retains a non-exclusive, paid-up, irrevocable, world-wide license to publish or reproduce the published form of this manuscript, or allow others to do so, for United States Government purposes.